\documentclass[twocolumn,preprintnumbers,amsmath,amssymb,superscriptaddress,nofootinbib,longbibliography, prb]{revtex4-2}

\usepackage{graphicx}% Include figure files
\usepackage{bm}% bold math
\usepackage{dsfont}
\usepackage[usenames,dvipsnames]{xcolor}
\usepackage{pstricks}
\usepackage[tight]{subfigure}
\usepackage{verbatim}
\usepackage{units}
\usepackage{multirow}
\usepackage{mathrsfs}
\usepackage{leftidx}
\usepackage{xspace}
\usepackage{diffcoeff}  %\dl x -> dx, \diff[n]{y}{x}[x_0] -> \frac{d^n y}{dx^n}|_{x_0}, \diff*{y}{x} -> \frac{d^n}{dx^n}y, \diffp for partial derivatives
\usepackage{bbm}
\usepackage{amsfonts,amssymb,amsmath}
\usepackage{mathtools}

\usepackage{array} 
\usepackage{tabularx}
\usepackage[normalem]{ulem}
\usepackage{bbold}
\usepackage{pifont}
\usepackage{hyperref}
\hypersetup{colorlinks=true,linktoc=all,linkcolor=blue,breaklinks=true,citecolor=blue,urlcolor=blue}

\usepackage[utf8]{inputenc}
\usepackage[english]{babel}

\definecolor{darkgreen}{RGB}{6, 153, 38}

\addto\captionsenglish{\renewcommand{\figurename}{FIG.}}

\AtBeginDocument{%
    \newwrite\bibnotes
    \def\bibnotesext{Notes.bib}
    \immediate\openout\bibnotes=\jobname\bibnotesext
    \immediate\write\bibnotes{@CONTROL{REVTEX42Control}}
    \immediate\write\bibnotes{@CONTROL{%
    apsrev42Control,author="08",editor="1",pages="1",title="0",year="1"}}
     \if@filesw
     \immediate\write\@auxout{\string\citation{apsrev42Control}}%
    \fi
}%

\newcommand{\beq}{\begin{equation}}
\newcommand{\eeq}{\end{equation}}

\DeclarePairedDelimiter{\abs}{\lvert}{\rvert}

\DeclarePairedDelimiterX\braket[2]{\langle}{\rangle}{#1\,\delimsize\vert\,\mathopen{}#2}
\DeclarePairedDelimiterX\ketbra[2]{\lvert}{\rvert}{#1\,\delimsize\rangle\mathopen{}\delimsize\langle\,\mathopen{}#2}

\DeclarePairedDelimiterX\Braket[2]{(}{)}{#1\,\delimsize\vert\,\mathopen{}#2}
\DeclarePairedDelimiterX\Ketbra[2]{\lvert}{\rvert}{#1\,\delimsize)\mathopen{}\delimsize(\,\mathopen{}#2}

\newcommand{\kB}{k_\text{B}}
\newcommand{\kBT}{k_\text{B}T}

\begin{document}

\title{
Scattering theory of thermal and bipolar thermoelectric diodes
}
\author{Jos\'e Balduque}
\affiliation{Departamento de F\'isica Te\'orica de la Materia Condensada, Universidad Aut\'onoma de Madrid, 28049 Madrid, Spain\looseness=-1}
\affiliation{Condensed Matter Physics Center (IFIMAC), Universidad Aut\'onoma de Madrid, 28049 Madrid, Spain\looseness=-1}
\author{Rafael S\'anchez}
\affiliation{Departamento de F\'isica Te\'orica de la Materia Condensada, Universidad Aut\'onoma de Madrid, 28049 Madrid, Spain\looseness=-1}
\affiliation{Condensed Matter Physics Center (IFIMAC), Universidad Aut\'onoma de Madrid, 28049 Madrid, Spain\looseness=-1}
\affiliation{Instituto Nicol\'as Cabrera (INC), Universidad Aut\'onoma de Madrid, 28049 Madrid, Spain\looseness=-1}
\date{\today}

\begin{abstract}
We investigate the minimal requirements that induce a nonreciprocal response to temperature differences in a mesoscopic electronic conductor. We identify two distinct mechanisms involved in electron-electron interactions, namely inelastic scattering and screening, that locally affect the internal properties of the device, leading to thermal and thermoelectric rectification effects in the absence of inversion symmetry. We propose resonant tunneling samples to efficiently exploit these effects, and find configurations acting as bipolar thermoelectric diodes whose current flows in the same direction irrespective of the sign of the temperature difference, a case of antireciprocity. 
\end{abstract}

\maketitle

%%%%%%%%%%%%%%%%%%%%%%%%%%%%%%%%%%%%%%%%%%%%%%%%%%%%%%%%%%%%%%%%%%%%%%%%%%%%%%
%%%%%%%%%%%%%%%%%%%%%%%%%%%%%%%%%%%%%%%%%%%%%%%%%%%%%%%%%%%%%%%%%%%%%%%%%%%%%%

\begin{figure}[b]
\includegraphics[width=\linewidth]{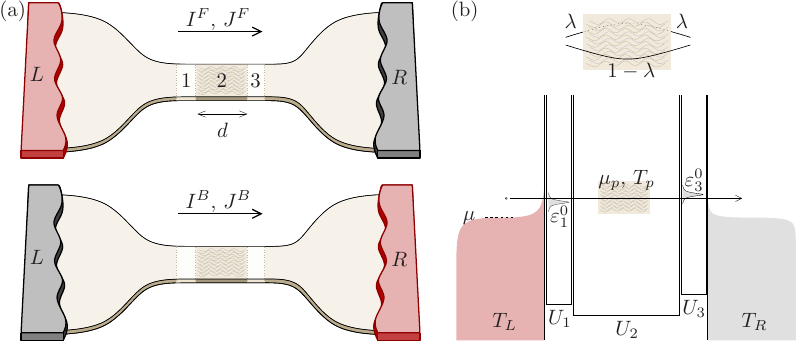}
\caption{\label{fig:scheme}
Scheme of a temperature-induced diode. (a) It consists on three regions: 1 and 3, where transport is elastic, and 2, where electrons may thermalize. Particle ($I$) and heat ($J$) currents responding to differences in the temperature $T_j$ of terminals $j=L,R$ are sensitive to the forward (F) or backward (B) configuration where temperature is increased in terminal L or R. (b) Scheme of a probe modelling thermalization in region 2 which absorbs (reflects) electrons with probability $\lambda$ ($1{-}\lambda$), and (c) of a configuration where regions 1 and 2 are tunnel junctions with resonances at energy $\varepsilon_\alpha^0$ with respect to the corresponding band bottom, $U_\alpha$.
%with respect to the overall equilibrium potential $U_0=0$.
}
\end{figure}

{\it Introduction}.---Macroscopic materials and devices typically show a linear response to temperature differences~\cite{ashcroft_book}, with deviations being described recently in extended low-dimensional systems~\cite{wang_quantum_2022,yamaguchi_microscopic_2024,arisawa_observation_2024,chatterjee_quasiparticle_2024}. This regime is governed by the principle of microreversibility, that has profound consequences such as Onsager reciprocity relations~\cite{onsager1931I,*onsager1931II}. This means that reversing the temperature differences applied to the two terminals of a conductor changes the sign of the charge and heat currents, but not their magnitude. In the quantum regime, and in the absence of strong interactions, these properties can be derived from the scattering matrix defining the conductor close to equilibrium~\cite{jacquod_onsager_2012}. 
Nanoscale devices~\cite{pekola_colloquium_2021} which avoid these constrains are strongly demanded  to define diodes whose conduction properties are sensitive to nonequilibrium states onchip (due to e.g., undesired hotspots or leakage heat flows): nonreciprocal currents are different in the forward (F) and backward (B) configurations, when a temperature increase $\Delta T$ holds either on the left or on the right terminal, see Fig.~\ref{fig:scheme}(a). Most proposals so far are based on specific realizations dominated by electron-electron interactions~\cite{xueOu_thermal_2008,kuo_thermoelectric_2010,
ruokola:2011,sierra_strongly_2014,sierra_nonlinear_2015,Vannucci2015,alexCarmen,tesser_heat_2022}, with so far very few experiments detecting nonreciprocal thermal or thermoelectric responses to temperature differences~\cite{scheibner_quantum_2008,svensson_nonlinear_2013,fast_geometric_2023}. Some other works can be interpreted in terms of noninteracting electrons, however requiring the coupling to additional degrees of freedom (phonons, photons)~\cite{segal_single_2008,martinez-perez2013,martinezPerez_rectification_2015,jiang:2015,guillem,craven_electron_2018,lu_quantum_2019,donald,cao_quantum_2024,mehring_hysteresis_2024}, a quantum detector~\cite{bredol_decoherence_2021,ferreira_exact_2023} or a third terminal~\cite{chiraldiode,
sanchez_single_2017,genevieveqpc,extrinsic,tesser_heat_2022}, all involving heat being dissipated elsewhere, rather than rectified. 
Photonic thermal rectification in quantum information systems is also intensively investigated~\cite{segal_spin_2005,ojanen_selection_2009,fratini_fabry_2014,schaller_collective_2016,ordonezMiranda_quantum_2017,barzanjeh_manipulating_2018,senior_heat_2020,bhandari_thermal_2021,iorio_photonic_2021,upadhyay_microwave_2024,poulsen_heatbased_2024,alfredo_photon}. However, an overall description of the microscopic origin of temperature-driven electronic diodes is missing which is not restricted to near equilibrium situations~\cite{sanchez_scattering_2013,whitney_nonlinear_2013,whitney_thermodynamic_2013,meair_scattering_2013,lopez_nonlinear_2013}.

Here we investigate the onset of temperature-driven diode effects by exploring the geometric and dynamic consequences of reduced dimensionality conductors: on one hand, they have a low capacitance~\cite{ihn_semiconductor_2009}; on the other hand, their size can be comparable to the carrier thermalization length~\cite{benenti:2017}. 
Both features emphasize the importance of electron-electron interactions: charge accumulation in the conductor alters the internal potential, while the momentum transfer involved in the Coulomb interaction results in inelastic scattering, phase randomization and thermalization within the nanostructure. 
Importantly, we will not require any external environment, such that energy is conserved in the conductor: at low temperatures, the electron-phonon coupling is negligible. The key aspect is that the state of the quantum system is sensitive to the temperature distribution in the two terminals, for which one additionally needs to impose broken (left-right) inversion symmetry. 

%%%%%%%%%%%%%%%%%%%%%%%%%%%%%%
%%%%%%%%%%%%%%%%%%%%%%%%%%%%%%
To keep the discussion simple, we use a scattering theory based description, which is well established close to the linear regime as long as electron-electron interactions can be treated in a mean field level~\cite{landauer_spatial_1957,landauer_conductance_1989,buttiker_admittance_1997}. Nontrivial extensions of the theory are needed that account for nonlinearities~\cite{christen_gauge_1996} (relating charge accumulation in the conductor to rectification and diode effects~\cite{buttiker_admittance_1997}) and inelastic scattering~\cite{engquist_definition_1981,buttiker:1986,buttiker:1988} (treated phenomenologically). 
For this, we consider the configuration sketched in Fig.~\ref{fig:scheme}(a): thermoelectric particle, $I$, and heat, $J$, currents through a quantum conductor are due to temperature differences between terminals L and R~\cite{sivan:1986,butcher:1990}. 
The conductor is partitioned in three, with regions $\alpha=1,3$ sandwiching region 2, of length $d$, where electrons are ballistic but have a finite probability $\lambda$ to thermalize. 
Despite its simplicity, this model can be applied to a wide range of configurations, including quantum wires~\cite{streda:1989,molenkamp:1990,guttman_thermopower_1995,kheradsoud:2019,haack:2019}, molecular junctions~\cite{paulsson:2003,reddy:2007}, chaotic cavities~\cite{sanchez:2011} (which enhance thermalization~\cite{rangi_disorder_2024}) or quantum dots~\cite{jaliel:2019}.
Remarkably, it not only allows us to identify the relevant mechanisms for thermoelectric and thermal rectification, it also predicts an antireciprocal response: a bipolar thermoelectric effect in particular configurations for which particles flow in the same direction independently of which terminal is hot. We discuss how this effect depends on the breaking of the microreversibility principle for the different involved mechanisms.

{\it Scattering theory}.--- The transport properties of a single-channel conductor are described by its scattering matrix, ${\cal S}(E)$. With the current density in terminal $j$
\begin{align}
\label{eq:currdens}
{\cal I}_j(E)=\frac{2}{h}\sum_k|{\cal S}_{kj}(E)|^2[f_j(E)-f_k(E)]
\end{align}
one writes the particle, $I_j={\int}{dE}{\cal I}_j(E)$, and heat currents, $J_j={\int}{dE}(E-\mu_j){\cal I}_j(E)$~\cite{sivan:1986}, where $f_j(E)=1/\{1+\exp[(E-\mu_j)/\kBT_j]\}$ is the Fermi function of terminal $j$ at temperature $T_j$ and electrochemical potential $\mu_j$, $h$ and $\kB$ are the Planck and Boltzmann constants, and the factor 2 accounts for spin degeneracy~\cite{heikkila_book}.
We consider the scattering region to have a piecewise uniform band bottom $\mathbf{U}\equiv\{U_\alpha\}$, which will influence the shape of ${\cal S}(E)={\cal S}(E,\mathbf{U})$. For later convenience, we split them into equilibrium (including the effect of gate voltages) and nonequilibrium contributions: $U_\alpha=U_\alpha^{\rm eq}+U_\alpha^{\rm neq}$.
Note that since the particle flow relies on the thermoelectric effect, ${\cal S}(E)$ needs to be energy dependent~\cite{benenti:2017}. 

We are interested in two-terminal conductors, with $j=L,R$, only driven by a temperature difference $\Delta T$ applied either to L (the forward) or to R (the backward case), with the opposite terminal being at temperature $T$, and both having the same electrochemical potential, $\mu=\mu_L=\mu_R$. The diode effect appears when the particle or heat currents, $X^i(T_L,T_R)=I_L^i,J_L^i$ ($i=F,B$), are nonreciprocal i.e., $X^F\neq-X^B$, with $X^F\equiv X(T+\Delta T,T)$ and $X^B\equiv X(T,T+\Delta T)$. 
We quantify this effect with the thermoelectric, $\mathcal{R}_I$, and thermal, $\mathcal{R}_J$, rectification coefficients~\cite{khandelwal_characterizing_2023}
\begin{equation}
    \mathcal{R}_X=\dfrac{X^F+X^B}{\abs{X^F}+\abs{X^B}},
\end{equation}
which saturate to $\pm1$ only when one of the currents is zero (perfect diode) or when both have the same sign. In the latest case, the diode is bipolar.

We readily see that assuming fully-noninteracting particles, for which the scattering matrix is independent of the reservoir temperatures, both currents $X^i$ are antisymmetric under the exchange $T_L\leftrightarrow T_R$, c.f. Eq.~\eqref{eq:currdens} for $j=L$ and $k=R$, resulting in no rectification, ${\cal R}_X=0$. A diode hence needs that the nanostructure is sensitive to the terminal temperatures.

In what follows, we explore the role of interactions as a requisite for thermal and thermoelectric rectification. We consider two separate ways in which interactions can modify the electron propagation through the system, namely inelastic scattering and screening effects.

%%%%%%%%%%%%%%%%%%%%%%%%%%%%%%%%%
\textit{Inelastic scattering}.--- When two electrons interact, they exchange momentum, involving that they change their energies and randomize their kinetic phases, limiting both the elastic and phase-coherent transport implicit in Eq.~\eqref{eq:currdens}, while conserving their total energy. 
In scattering theory, this effect is routinely described phenomenologically by introducing a fictitious probe~\cite{buttiker:1986,buttiker:1988,dAmato_conductance_1990,dejong_semiclassical_1996,Datta1995,bedkihal_probe_2013,utsumi_fluctuation_2014} which absorbs electrons with probability $\lambda$~\cite{SM} and reinjects them with a random phase and thermalized with a distribution $f_p(E)$ whose electrochemical potential, $\mu_p$, and temperature, $T_p$, are determined by imposing that the probe injects no particle and no heat currents on average, $I_p=J_p=0$. 
Intuitively $\lambda$ relates the inelastic scattering length $l_{inel}$ to the typical size of the system, $l$: $\lambda\ll1$ when $l_{inel}\gg l$, and $\lambda\approx1$ in the opposite limit. 
Quantum Hall realizations relate $\lambda$ to the transmission of a quantum point contact~\cite{roulleau_tuning_2009}.
Trajectories involving the probe do not obey microreversibility. On a microscopic description of the particular models, there may be small deviations from a Fermi distribution function, specially at very low temperatures and far from equilibrium~\cite{jauho_book} which are not relevant for our discussion here. Our thermalization probe, sketched in Fig.~\ref{fig:scheme}(b), is hence formally equivalent to a thermometer~\cite{bergfield_probing_2013,shastry_temperature_2016,zhang_local_2019,shastry_scanning_2020} and useful to discuss thermalization in hot-carrier solar cells~\cite{tesser_thermodynamic_2023},  broken-time-reversal-induced electrical diodes~\cite{bedkihal_probe_2013} and correlations in edge channels~\cite{braggio_nonlocal_2024}.
Note that while the flow of heat will be determined by $T_L-T_R$,  the particle currents may be affected by the competition of the thermoelectric effect and the developed $\mu_p-\mu$.

The rectification properties of thermalization is proven analytically by considering a simple configuration in which all electrons in the conductor are thermalized ($\lambda=1$), with region 3 being transparent (so ${\cal S}_{pR}=\mathbb{1}$) and region 1 being a barrier with transmission probability ${\cal T}(E)=|{\cal S}_{pL}(E)|^2$, c.f. Fig.~\ref{fig:scheme}(a). 
In the limit $\mu{-}U_\alpha\gg\kBT$, the contribution of terminal R to the currents into the probe are given by ${\int}{dE}[f_R(E)-f_p(E)]=\mu-\mu_p$ and ${\int}{dE}E[f_R(E)-f_p(E)]=\pi^2(T_R^2-T_p^2)/6$. The other contribution to the probe currents coincides with $X^i$, which are then determined by the probe conditions:
\begin{align}
I^i=\frac{2}{h}(\mu_{p,i}-\mu)\quad{\rm and}
\quad J^i=\frac{\pi^2}{3h}(T_{p,i}^2-T_{R,i}^2),
\end{align}
with $T_{R,F}=T$ and $T_{R,B}=T{+}\Delta T$. The spatial asymmetry here is due to the position of the thermalization region with respect to the thermoelectric element, which is enough to have $T_p^F\neq T_p^B$. A thermoelectric diode needs that ${\cal T}(E)$ breaks electron-hole symmetry, so region 1 has a finite electric response giving $\mu_p^F\neq\mu_p^B$~\cite{SM}.

\begin{figure}[t]
    \includegraphics[width=\linewidth]{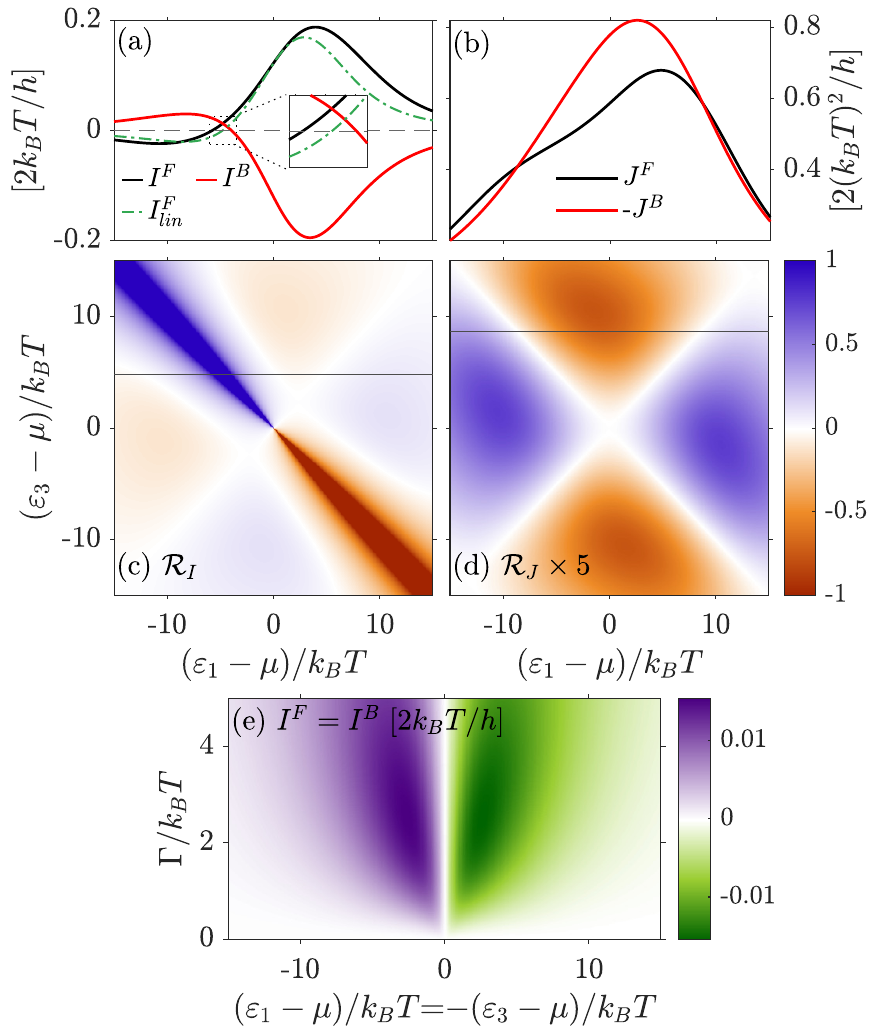}
    \caption{\label{fig:Th only currents}Thermalization diode. Forward and backward (a) particle and (b) heat currents as functions of $\varepsilon_1$ tuned along $(\varepsilon_3{-}\mu)/\kBT=4.8$ and $8.7$, respectively, indicated by black lines  in (c) and (d). The later panels show the thermoelectric and heat rectification coefficients, as functions of both resonances. The thermalization probe is fully coupled ($\lambda=1$), and $\mu\gg U_\alpha$, avoiding band bottom effects. The linear response current, $I_{lin}^F=L^{(1)F}\Delta T$, is shown for comparison in (a).  Parameters: $\Gamma = 3\kBT$, $\mu=40\kBT$ and $\Delta T/T=1$. }
\end{figure}

We now consider a more specific setup, sketched in Fig.~\ref{fig:scheme}(c), which allows us to control the degree of asymmetry experimentally~\cite{prance:2009,jaliel:2019}. We choose the scattering regions $1$ and $3$ to be resonant-tunneling barriers (RTB), known efficient thermoelectric devices~\cite{staring_coulomb_1993,dzurak_thermoelectric_1997,nakpathomkun:2010,svensson_lineshape_2012,josefsson_quantum_2018}. We model them by Breit-Wigner resonances~\cite{breitwigner} with energy $\varepsilon_\alpha^0$ and inverse lifetime $\Gamma_\alpha/\hbar$~\cite{buttiker:1988} and transmission amplitude~\cite{SM}  
\begin{align}
\label{eq:Breit-Wigner}
%{\cal S}_{\alpha,12}{=}{\cal S}_{\alpha,21}{\equiv}
\tau_\alpha = \frac{-i\Gamma_\alpha}{E-\varepsilon_\alpha^0 -U_\alpha + i\Gamma_\alpha}, \quad (\alpha{=}1{,}3).
\end{align}
For simplicity, we assume $\Gamma_1{=}\Gamma_3{\equiv}\Gamma$, and a totally-coupled probe ($\lambda=1$) so direct elastic transport between terminals L and R is suppressed, a configuration of experimental relevance~\cite{jaliel:2019,nilsson_single_2016,barker_individually_2019}.
We incorporate $U_\alpha^{\rm eq}$, which allows us to control the system asymmetry via gate voltages, in the resonance energies $\varepsilon_\alpha=\varepsilon_\alpha^0{+}U_\alpha^{\rm eq}$, and neglect $U_{\alpha}^{\rm neq}$ here. 

The resulting currents and rectification coefficients are plotted in Figs.~\ref{fig:Th only currents}(a)-(d) by tuning the resonance energies of the barriers.
As expected, broken inversion symmetry when $\varepsilon_1\neq\varepsilon_3$ results in finite ${\cal R}_I$ and ${\cal R}_J$, while ${\cal R}_X=0$ for symmetric cases with $\varepsilon_1=\varepsilon_3$. 

The antisymmetric configuration with $\varepsilon_1-\mu=\mu-\varepsilon_3$ is particularly interesting: while inversion symmetry is maximally broken, it respects electron-hole symmetry in equilibrium. The two RTB have opposite thermoelectric contributions resulting in a vanishing linear response for particles, see Fig.~\ref{fig:Th only currents}(a): writing $I^i=\sum_n L^{(n),i}\Delta T^n$, we get $L^{(1),i}=0$, with the probe developing a temperature $T_p^F=T_p^B=T+\Delta T/2$ and electrochemical potentials of opposite sign in F and B, $\mu_p^F+\mu_p^B=2\mu$~\cite{SM}. This affects the nonlinear terms (of order $\geq\Delta T^2$), making $I^F$ and $I^B$ vanish at different points (determined by the corresponding $\mu_p$ and $T_p$) when, e.g., tuning $\varepsilon_1$, see inset in Fig.~\ref{fig:Th only currents}(a).
At the vanishing points, the current flows only in one of the configurations, as an ideal thermoelectric diode with ${\cal R}_I=1$ (however working only for a particular $\Delta T$). 
Most remarkably, between vanishing points, the current flows in the same direction irrespectively of which terminal is hot. This is what we call a bipolar thermoelectric (or antireciprocal) diode. Furthermore, at the antisymmetric condition $\varepsilon_1+\varepsilon_3=2\mu$, the problem symmetry imposes $I^F=I^B$~\cite{SM}, with their sign being tunable and maximal for widths $\Gamma\approx2\kBT$ (consistent with similar configurations~\cite{jordan:2013,szukiewicz_optimisation_2016,balduque_coherent_2024}), see Fig.~\ref{fig:Th only currents}(e). 

Differently, the heat currents do not change sign (as imposed by the second law), see Fig. \ref{fig:Th only currents}(b), which avoids a bipolar thermal diode if $\mu_j=\mu$. We find ${\cal R}_J\sim15\%$, cf. Fig.~\ref{fig:Th only currents}(d), and ${\cal R}_J=0$ for $(\varepsilon_1-\mu)=\pm(\varepsilon_3-\mu)$, where both RTBs conduct heat equally.

\textit{Screening effects}.--- The injection of a charge current is able to alter the potential landscape of the conductor via charge accumulation and screening effects in the nearby gates defining the scattering regions~\cite{levinson_potential_1989,*levinson1989,buttiker_capacitance_1993}. This consequence of electron-electron interaction is in fact needed to ensure gauge invariance in nonlinear scenarios~\cite{christen_gauge_1996}. Deviations from the linear regime leading to finite thermal and thermoelectric rectification have been discussed in the weakly interacting regime~\cite{sanchez_scattering_2013,whitney_thermodynamic_2013,meair_scattering_2013,lopez_nonlinear_2013}. 

We again test this mechanism in the setup of Fig.~\ref{fig:scheme}(c), now  with $\lambda=0$, so that transport between terminals is fully elastic. Then, an important contribution will be the internal reflections in region 2, leading to Fabry-Perot interferences. Scattering at the barriers is again given by Eq.~\eqref{eq:Breit-Wigner}.
Screening affects the internal electrostatic energies, $\mathbf{U}$, by developing finite $U^{\rm neq}_\alpha(T_L,T_R)$, and modifies the scattering matrix of the whole system, ${\cal S}(E,\mathbf{U})$~\cite{SM}.
The energies $\mathbf{U}$ are linked by 
\begin{align}
\label{eq:dq_C}
    \delta q_\alpha = \sum_{\beta=1,2,3 \neq \alpha} C_{\alpha \beta}(U_\beta-U_\alpha)/(-e),
\end{align}
where $\delta q_\alpha$ is the injected charge in region $\alpha$ with respect to the equilibrium scenario, and $C_{\alpha \beta}$ the geometric capacitance between regions $\alpha$ and $\beta$, treated here as a parameter. The excess charge can be calculated with the aid of the partial local densities of states of the system, or injectivities~\cite{buttiker_capacitance_1993}, 
\begin{equation}
    \nu_{j\alpha}(E,\mathbf{U})=\dfrac{i}{4\pi}\sum_{k}\left[{\cal S}^{\dagger}_{kj}\dfrac{\delta {\cal S}_{kj}}{\delta U_\alpha}-\dfrac{\delta {\cal S}^{\dagger}_{kj}}{\delta U_\alpha}{\cal S}_{kj}\right],
\end{equation}
defining the overlap of waves in region $\alpha$ with states injected at terminal $j$, via
\begin{align}
\label{eq:dq_q}
%   \frac{\delta q_\alpha}{-e} = {\int}dE \sum_j  [\nu_{j\alpha}(E,\mathbf{U})f_j(E){-}\nu_{j\alpha}(E,\mathbf{U}^{\rm eq})f_{\rm eq}(E)],
   \frac{\delta q_\alpha}{-e} = {\int}dE \sum_j  [\nu_{j\alpha}(E,\mathbf{U})f_j(E){-}\nu_{j\alpha}^{\rm eq}f_{\rm eq}(E)],
\end{align}
with $\nu_{j\alpha}^{\rm eq}$ and $f_{\rm eq}(E)$ evaluated at equilibrium. Note that we do not assume a weak deviation from equilibrium~\cite{sothmann_nonlinear_2019}.

\begin{figure}[t]
    \includegraphics[width=\linewidth]{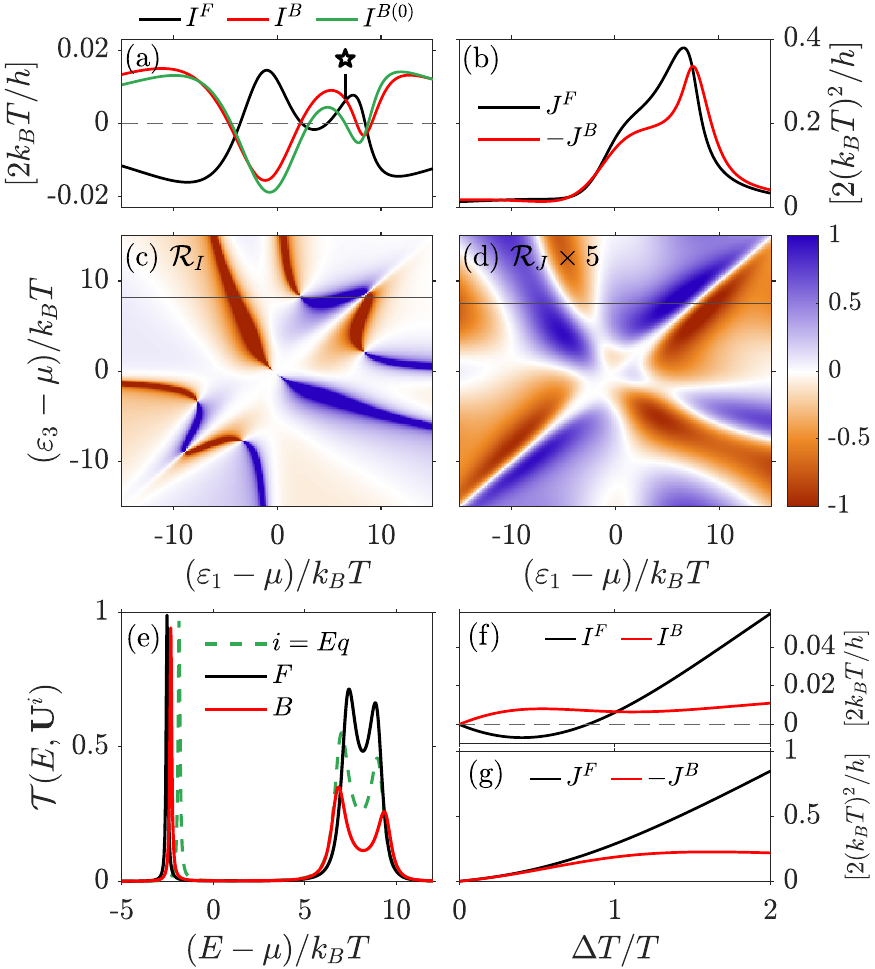}  \caption{\label{fig:rectification_NL_gamma1_x2}Rectification by screening. (a) Particle and (b) heat currents as functions of $\varepsilon_1$, for fixed $(\varepsilon_3-\mu)/\kBT=8.2$ and $7.5$ as marked by black lines in (c) and (d), respectively. The later show the dependence of ${\cal R}_I$ and ${\cal R}_J$ with gating the RTBs. The green curve in (a) shows the current obtained by imposing the equilibrium transmission probability, for comparison. (e) Transmission probability in equilibrium and in the F and B configurations at the point marked by $\star$ in (a), for which the temperature dependence of the charge and heat currents are shown in (f) and (g). Parameters: $\lambda{=}0$, $d=2 h/\sqrt{8m\kBT}$, $\Gamma = \kBT$, $\mu=40\kBT$, $\Delta T/T=1$. }
\end{figure}

We show in Figs.~\ref{fig:rectification_NL_gamma1_x2}(a) and \ref{fig:rectification_NL_gamma1_x2}(b) the resulting currents in the limit $C_{\alpha\beta}\to0$ for small conductors. The thermoelectric response shows an oscillating behaviour, with sign changes close to the crossings of the barrier resonances by the chemical potential, and at the crossing of the two resonances. As $\mathbf{U}$ depends on the temperature distribution, these crossings occur at different points in F and B, again resulting in regions with ${\cal R}_I=1$, see Fig.~\ref{fig:rectification_NL_gamma1_x2}(c). The heat currents do not change sign, finding wide regions with strong rectification (${\cal R}_J\sim0.2$) close to the resonance condition $\varepsilon_1\approx\varepsilon_3$, see Fig.~\ref{fig:rectification_NL_gamma1_x2}(d).

Screening hence also induces a bipolar thermoelectric diode, in this case respecting microreversibility: $|{\cal S}_{LR}(E,\mathbf{U})|^2=|{\cal S}_{RL}(E,\mathbf{U})|^2={\cal T}(E)$. To further understand this mechanism, let us concentrate on the particular configuration marked by $\star$ in Fig.~\ref{fig:rectification_NL_gamma1_x2}(a), where $I^F{=}I^B$. 
There, both resonances are over the chemical potential and slightly detuned, $\varepsilon_{3}>\varepsilon_1>\mu$. 
The transmission probability ${\cal T}(E)$ [see Fig.~\ref{fig:rectification_NL_gamma1_x2}(e)] shows two main features with opposite thermoelectric contributions: a wide double peak around $\varepsilon_1,\varepsilon_3>\mu$, and a sharp peak at $E<\mu$ due to the Fabry-Perot interference in $\alpha$=2. 
Note that ${\cal T}(E\approx\mu)$ is flat, which suppresses the linear term ($L^{(1)}{\propto}\partial{\cal T}/\partial E$) according to the Mott formula~\cite{benenti:2017}.   
In F, increasing $T_L$ increases $\delta q_1$ and hence $U_1$, bringing the two Breit-Wigner resonances closer and resulting in a sharper and higher double peak at positive energies, see Fig.~\ref{fig:rectification_NL_gamma1_x2}(e). Oppositely, the corresponding increase of $U_3$ in B separates the two resonances, making the double peak wider and lower, i.e., reducing its contribution to $I^B$. Eventually, the opposite contribution of the Fabry-Perot peak at negative energies dominates, inducing $I^B$ to change sign~\cite{SM}.
This bipolar diode is hence induced not only by screening but also by quantum interference. 

The bipolar effect persists for higher $\Delta T$, cf. Fig.~\ref{fig:rectification_NL_gamma1_x2}(f), and the thermal diode is robust: one polarity exhibits vanishing and even negative thermal differential conductance for large $\Delta T$~\cite{SM}, see Fig.~\ref{fig:rectification_NL_gamma1_x2}(g). We attribute this effect to the same mechanism discussed above.  

\begin{figure}[t]
    \includegraphics[width=\linewidth]{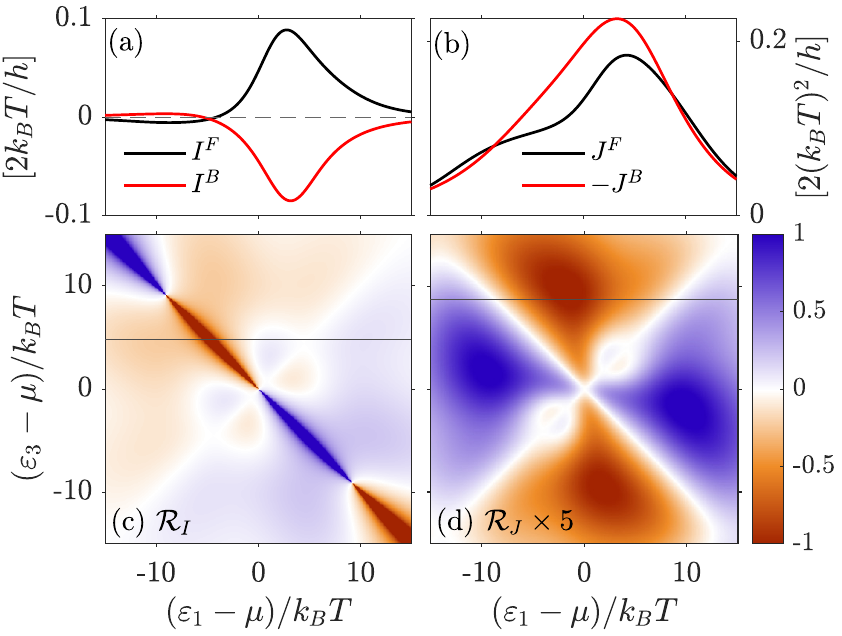}
    \caption{\label{fig:mixed}Mixed configuration. Same as Figs.~\ref{fig:rectification_NL_gamma1_x2}(a)-(d) for $\lambda{=}1$. 
    }
\end{figure}

{\it Mixed regime}---. To confirm the quantum interference origin of the screening-induced bipolar thermoelectric diode, we compute a mixed configuration where screening coexists with thermalization, cf. Fig.~\ref{fig:mixed}. For $\lambda=1$, the probe avoids any Fabry-Perot interference in region 2. Hence screening only contributes to shift $U_1$ and $U_3$. By comparing Figs.~\ref{fig:Th only currents}, \ref{fig:rectification_NL_gamma1_x2} and \ref{fig:mixed} we see that the currents are indeed dominated by thermalization, and the interference-induced off-diagonal features in ${\cal R}_I$ are strongly suppressed~\cite{SM}. Still, we find a thermalization-induced bipolar effect around $\varepsilon_1-\mu=\mu-\varepsilon_3$.

%%%%%%%%%%%%%%%%%%%%%%%%%%%%%%%%%%%%%%%%%%%%%%%%%%%%%%%%%%%%%%%%%%%%%%%%%%%%%%
{\it Conclusions}.--- We have provided a fully quantum mechanical description of temperature-induced mesoscopic diodes by considering interaction-induced thermalization and screening affecting elastic transport (however strongly correlated effects like Coulomb blockade are not covered~\cite{kouwenhoven:1997}). With this we find diode effects both for particle and thermal currents. Remarkably, we find an {\it antireciprocal} response in the form of a bipolar thermoelectric effect caused by both thermalization and screening-controlled quantum interference in ballistic resonant tunneling configurations.
While the fully thermalized case can be interpreted in terms of locally reversing currents between three regions (the reservoirs and the internal thermalization region) at different thermal configurations, the fully coherent case exploits nonequilibrium states and quantum interference to achieve antireciprocity while maintaining microreversibility. 
In both cases, it occurs when sharp spectral features with opposite but similar in magnitude thermoelectric contributions depend differently on the local reaction of the conductor to the nonequilibrium situation. We find measurable currents of a few nA at mK temperatures in state of the art configurations~\cite{jaliel:2019,barker_individually_2019}. 

Related effects have been predicted before however requiring dissipation~\cite{lu_quantum_2019,bredol_decoherence_2021,ferreira_exact_2023} or multiterminal configurations~\cite{sanchez_single_2017,genevieveqpc,extrinsic}. Differently, our treatment imposes particle and energy conservation in the two-terminal diode. Our results add to other recently found bipolar nonlinear thermoelectric functionalities in superconducting junctions~\cite{marchegiani_nonlinear_2020,germanese_bipolar_2022,battisti_bipolar_2024} (based on spontaneous symmetry breaking mechanisms), contributing to set the base for thermally-driven onchip devices.

%%%%%%%%%%%%%%%%%%%%%%%%%%%%%%%%%%%%%%%
%\acknowledgements
We thank Y. Tokura and A. Braggio for useful discussions.
We acknowledge funding from the Spanish Ministerio de Ciencia e Innovaci\'on via grants No. PID2019-110125GB-I00 and No. PID2022-142911NB-I00, and through the ``Mar\'{i}a de Maeztu'' Programme for Units of Excellence in R{\&}D CEX2023-001316-M.

%%%%%%%%%%%%%%%%%%%
%%%%%%%%%%%%%%%%%%%
%%%%%%%%%%%%%%%%%%%
%%%%%%%%%%%%%%%%%%%
%%%%%%%%%%%%%%%%%%%
%%%%%%%%%%%%%%%%%%%
%%%%%%%%%%%%%%%%%%%
%%%%%%%%%%%%%%%%%%%
%%%%%%%%%%%%%%%%%%%
%%%%%%%%%%%%%%%%%%%

%%%%%%%%%%%%%%%%%%%%%%%%%%%%%%%%%%%%%%%%%%%%%%%%%%%%%%%%%%%%%%%%%%%%%%%%%%%%%%

\bibliography{biblio.bib}
%%%%%%%%%%%%%%%%%%%%%%%%%%%%%%%%%%%%%%%

%%%%%%%%%%%%%%%%%%%%%%%%%%%%%%%%%%%%%%%

\newpage

\setcounter{equation}{0}
\renewcommand{\theequation}{S\arabic{equation}}

\setcounter{figure}{0}
\renewcommand{\thefigure}{S\arabic{figure}}

\onecolumngrid
\vspace{\columnsep}
\section*{Supplementary material for: 
``Scattering theory of thermal and bipolar thermoelectric diodes"}
\vspace{\columnsep}
\twocolumngrid

%begin{document}

%%%%%%%%%%%%%%%%%%%%%%%%%%%%%%%%%%%%%%%%%%%%%%%%%%%%%%%%%%%%%%%%%%%%%%%%%%%%%%
%%%%%%%%%%%%%%%%%%%%%%%%%%%%%%%%%%%%%%%%%%%%%%%%%%%%%%%%%%%%%%%%%%%%%%%%%%%%%%

In this supplementary material we give additional details useful for the discussion in the main text. Section,~\ref{sec:SMscattering} presents the scattering matrices of the different regions of the conductor and how they lead to the global scattering matrix of the system. Section~\ref{sec:minimal} gives a derivation of the analytical results demonstrating the rectification induced by a thermalization probe in a minimal model. The model including two resonant tunneling barriers separated by a thermalization probe is analyzed in Secs.~\ref{Ssec:linear} (linear response) and \ref{Ssec:bipolarRect}. The later includes a discussion of the physical mechanism leading to the bipolar thermoelectric diode, the effect of the coupling to the thermalization probe and the temperature dependence. The physical interpretation and temperature dependence of the screening-induced bipolar diode is given in Sec.~\ref{Ssec:bipolarRect_screening}. Finally, we include in Sec.~\ref{Ssec:currents} plots of the currents and rectification coefficients in different parameter configurations. 

\section{Scattering matrices}
\label{sec:SMscattering}

%Here we give details on the considered scattering matrices.

\subsection{Matrix for the coupling to the probe}

The coupling to the probe terminal is introduced by the scattering matrix~\cite{buttiker:1988}: 
\begin{eqnarray}\label{Sdeph}
\displaystyle
{
\mathcal{S}^\lambda=
%(s_{\rm u},s_{\rm d})=
\left(\begin{array}{cccc}
0 & \sqrt{1{-}\lambda} & 0 & i\sqrt{\lambda}\\ 
\sqrt{1{-}\lambda} & 0 & i\sqrt{\lambda} & 0\\
i\sqrt{\lambda} & 0 & \sqrt{1{-}\lambda} & 0\\
0 & i\sqrt{\lambda} & 0 & \sqrt{1{-}\lambda}\\
\end{array}  \right),
}
\end{eqnarray}
where $\lambda$ is the probability of an electron in region 2 to be absorbed and thermalized by the probe.  In the matrix, indices 1 and 2 correspond to the conductor channels connecting to regions 1 and 3; indices 3 and 4 correspond to channels connected to the probe terminal. Note that $\lambda$ is treated in the same footing as other transmission probabilities in the scattering region.

In the configuration considered in the main text, for $\lambda=0$, transport between the other conductor terminals (L and R) is elastic. For $\lambda=1$, all electrons are thermalized to the probe distribution when entering the central region. Transport can in that case be understood in terms of scattering between three thermal reservoirs (L, R and the probe) separated by regions 1 and 3.

\subsection{Resonant tuneling barriers}

In a mesoscopic conductor, resonant tunneling appears typically in double-barrier structures forming quantum dots or quantum wells. Considering symmetric barriers and a single resonant state, the single-channel scattering can be described by a Breit-Wigner~\cite{breitwigner} scattering matrix 
\begin{eqnarray}\label{Sres}
\displaystyle
{
\mathcal{S}^{\rm RTB}=
%(s_{\rm u},s_{\rm d})=
\left(\begin{array}{cc}
1+\tau & \tau\\ 
\tau & 1+\tau\\ 
\end{array}  \right),
}
\end{eqnarray}
with transmission amplitude
\begin{align}
\label{eq:Breit-Wigner}
\tau = \frac{-i\Gamma}{E-\varepsilon + i\Gamma}, 
\end{align}
where $\varepsilon$ is the energy of the resonant state and $\Gamma/\hbar$ is its inverse lifetime due to coupling to the modes outside the double barrier~\cite{buttiker:1988}. 

\subsection{Scattering matrix for the total system}

When the transport is fully elastic in the system the total scattering matrix have to account 
for processes that involve both scattering regions $\mathcal{S}_{1}$, $\mathcal{S}_{3}$, and the possible coherent reflections between them, where electrons accumulate a kinetic phase $k(E)d$; with $k(E)=\sqrt{2m(E-U_2)}$ the electron wavenumber in region $2$. This is done by taking into account that e.g., the right-outgoing wave from
region $1$ is the left-ingoing wave at region $3$, multiplied by a phase factor $e^{ik(E)d}$, and solving for all the outgoing waves as functions of the ingoing ones \cite{Datta1995,extrinsic}. Defining $E_\alpha=E-U_\alpha$, the total scattering matrix reads:

\begin{gather}
\begin{aligned}
\label{eq: S_nonlinear}
\mathcal{S}(E,\mathbf{U})&=   
\begin{pmatrix}
    r(E,\mathbf{U}) & \tau(E,\mathbf{U}) \\
     \tau(E,\mathbf{U}) & r'(E,\mathbf{U})
\end{pmatrix},\\
    \tau(E,\mathbf{U}) &= \dfrac{e^{ik(E)d}\Gamma^2}{e^{ik(E)d/l_0}E_1E_3+(E_1-i\Gamma)(E_3-i\Gamma)}\\
    r(E,\mathbf{U}) &= \dfrac{E_1(E_3-i\Gamma)-e^{ik(E)d}E_3(E_1-i\Gamma)}{e^{ik(E)d}E_1E_3+(E_1-i\Gamma)(E_3-i\Gamma)}\\
    r'(E,\mathbf{U})&=\dfrac{E_3(E_1-i\Gamma)-e^{ik(E)d}E_1(E_3-i\Gamma)}{e^{ik(E)d}E_1E_3+(E_1-i\Gamma)(E_3-i\Gamma)}.
\end{aligned}
\end{gather}

One should be careful when using the scattering matrix obtained by this approach to calculate injectivities for energies below $\mathbf{U}$, as it gives spurious divergences near the band bottom that are not present when using exact wave-function matching methods. This is relevant for the determination of the excess injected charge, as it involve the calculation of an integral starting from the energy origin. Here we have dealt numerically with this issue.

\section{Minimal model with thermalization}
\label{sec:minimal}

\begin{figure}[t]
    \includegraphics[width=\linewidth]{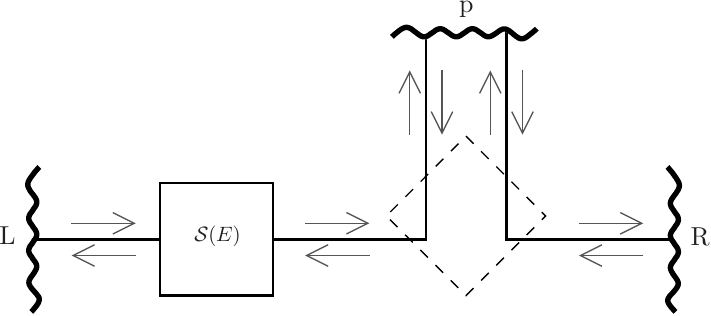}
    \caption{\label{fig:minimal}Minimal configuration showing rectification by thermalization with a fully transparent coupling ($\lambda=1$) to a probe terminal, p. A barrier represented by the scattering matrix ${\cal S}$ is placed between terminal L and the probe.  }
\end{figure}

Consider the situation depicted in Fig.~\ref{fig:minimal}, with a fully transparently coupled ($\lambda=1$) thermalization probe terminal, p, and a scattering region connecting it with terminal L with a transmission probability ${\cal T}(E)=|{\cal S}_{Lp}(E)|^2$. Electrons injected from terminal R are all absorbed by the probe. We impose $\mu_L=\mu_R=\mu$ to the conductor terminals and $I_p=J_p=0$ to the probe. 

The conductor currents are given by 
\begin{align}
\label{Seq:IL}
I_L&=\frac{2}{h}\int{dE}{\cal T}(E)(f_L-f_p)\\
\label{Seq:IR}
I_R&=\frac{2}{h}\int{dE}(f_R-f_p)=\frac{2}{h}(\mu-\mu_p),
\end{align}
for particles and 
\begin{align}
J_L&=\frac{2}{h}\int{dE}(E-\mu){\cal T}(E)(f_L-f_p)\\
J_R&=\frac{2}{h}\int{dE}(E-\mu)(f_R-f_p)=\frac{\pi^2}{3h}(T_R^2-T_p^2),
\end{align}
for heat. In the probe, $I_p=-I_L-I_R$ and $J_p=-J_L-J_R$. With charge and energy conservation, the current through the conductor is fully determined by the exact expressions for $I_R$ and $J_R$ above given by $\mu-\mu_p$ and $T_R^2-T_p^2$. In order to have a diode effect, such that $I_L^F\neq I_L^B$ and $J_L^F\neq J_L^B$, it is hence enough to prove that $\mu_p^F\neq\mu_p^B$ and $T_p^F\neq T_p^B$, respectively. 

For the thermoelectric case, it is easy to show that if ${\cal T}$ is energy independent, so it induces no thermoelectric effect in the probe, we get $I_L=2{\cal T}(\mu-\mu_p)/h$ independently of the temperature $T_L$. Therefore, the only current-conserving solution is $\mu_p=\mu$, indeed giving $I^F=I^B=0$. If we now consider an energy-dependent transmission probability, 
\begin{equation}
I_L^F=\frac{2}{h}\int{dE}{\cal T}(E)[f(T+\Delta T)-f_p]
\end{equation}
and \begin{equation}
I_L^B=\frac{2}{h}\int{dE}{\cal T}(E)[f(T)-f_p]
\end{equation}
are necessarily different, as long as the scatterer breaks electron-hole symmetry i.e., if it has a finite thermoelectic response. In that case, from Eq.~\eqref{Seq:IR} we get $\mu_p^F\neq\mu_p^B$. 

For the heat currents, it is sufficient to assume that $0<{\cal T}<1$. In the simplest case where it is energy independent, we get
\begin{gather}
\begin{aligned}
J_L^F&=\frac{{\cal T}\pi^2}{3h}[(T{+}\Delta T)^2-(T_p^F)^2]=\frac{\pi^2}{3h}[T^2-(T_p^F)^2]\\
J_L^B&=\frac{{\cal T}\pi^2}{3h}[T^2-(T_p^F)^2]=\frac{\pi^2}{3h}[(T{+}\Delta T)^2-(T_p^F)^2].
\end{aligned}
\end{gather}
Solving for the probe temperatures, we get:
\begin{align}
(T_p^F)^2&=({\cal T}-1)T^2+{\cal T}(\Delta T^2+2T\Delta T)\\
(T_p^B)^2&=({\cal T}-1)T^2-{\cal T}(\Delta T^2+2T\Delta T), 
\end{align}
clearly showing a diode effect, $T_p^F\neq T_p^B$.

\section{Linear response of a thermalization probe}
\label{Ssec:linear}

Consider two scattering regions, $\alpha=1,3$ separated by a transparently coupled probe terminal, such that all electrons injected from terminals L and R are either reflected at the barriers or absorbed by the probe. The electrochemical potential, $\mu_p$, and temperature, $T_p$, of the probe need to be calculated to fulfill the conditions $I_p=J_p=0$. Current is only due to a temperature difference $\Delta T$ applied to terminal $L$ in the forward (F), or to $R$ in the backward configuration (B). Let us consider first the forward case, where we expand the particle and heat currents as:
\begin{align}
I_L^F&=-G_1\Delta\mu_p^F+L_1(\Delta T-\Delta T_p^F)\\
I_R^F&=-G_3\Delta\mu_p^F-L_3\Delta T_p^F\\
J_L^F&=-M_1\Delta\mu_p^F+K_1(\Delta T-\Delta T_p^F)\\
J_R^F&=-M_3\Delta\mu_p^F-K_3\Delta T_p^F,
\end{align}
where $\Delta\mu_p^F=\mu_p^F-\mu$ and $\Delta T_p^F=T_p^F-T$. $G_\alpha$, $L_\alpha$, $M_\alpha$ and $K_\alpha$ are the electrical conductance, Seebeck coefficient, Peltier coefficient and thermal conductance of scattering region $\alpha$, respectively. Solving for the probe conditions, we get:
\begin{gather}
\begin{aligned}
\label{Seq:linearmupFTpF}
\Delta\mu_p^F=\frac{L_1K_3-L_3K_1}{G_\Sigma K_\Sigma-L_\Sigma M_\Sigma}\Delta T\\
\Delta T_p^F=\frac{G_\Sigma K_1-L_1 M_\Sigma}{G_\Sigma K_\Sigma-L_\Sigma M_\Sigma}\Delta T,
\end{aligned}
\end{gather}
where the subindex $\Sigma$ indicates the sum over L and R.
With these, we get the currents:
\begin{gather}
\begin{aligned}
\label{Seq:linearcurrF}
I_L^F&=\frac{G_1L_3K_1+G_3L_1K_3-L_1L_3M_\Sigma}{G_\Sigma K_\Sigma-L_\Sigma M_\Sigma}\Delta T\\
J_L^F&=\frac{K_1K_3G_\Sigma-K_1L_3M_3-K_3L_1M_1}{G_\Sigma K_\Sigma-L_\Sigma M_\Sigma}\Delta T.
\end{aligned}
\end{gather}
For the backward currents, we simply replace $1\leftrightarrow3$ in Eqs.~\eqref{Seq:linearmupFTpF} and \eqref{Seq:linearcurrF}. This way we obtain 
\begin{equation}
\Delta\mu_p^F=-\Delta\mu^B,
\end{equation}
as well as $I_L^F=-I_L^B$ and $J_L^F=-J_L^B$.
As expected, there is no rectification in the linear regime.

\subsection{Antisymmetric configuration}
\label{Ssec:antisymtherm}

A particularly interesting case is when the transmission of region 3, ${\cal T}_3(E)$ is the reflection of ${\cal T}_1(E)$ over the electrochemical potential. Then, $G_1=G_3$, $L_1=-L_3$, $M_1=-M_3$ and $K_1=K_3$. In that case, we get: 
\begin{equation}
\Delta\mu_p^F=\frac{L_1}{2G_1}\Delta T\quad{\rm and}\quad \Delta T_p^F=\frac{\Delta T}{2},
\end{equation}
and 
\begin{equation}
I_L^F=0\quad{\rm and}\quad J_L^F=\frac{K}{2}\Delta T.
\end{equation}
In the same way, $I_L^B=0$. 
Therefore in this case only the nonlinear contributions are responsible for the thermoelectric particle currents.

\section{Bipolar diode by thermalization}
\label{Ssec:bipolarRect}

\begin{figure}[t]
    \includegraphics[width=\linewidth]{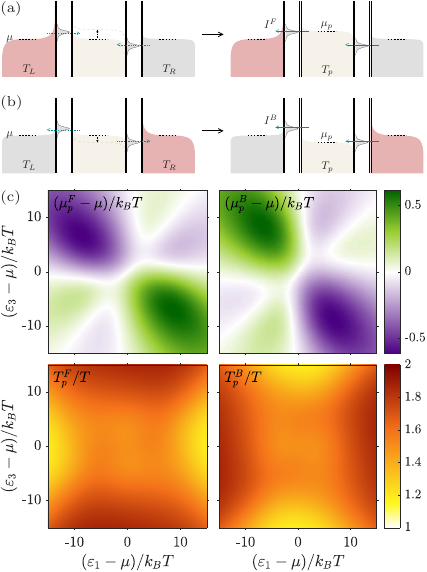}
    \caption{\label{fig:bipolarTh}Mechanism for bipolar rectification induced by the thermalization probe. The two panels show how the contribtions to the particle flows through the two barriers change with the chemical potential developed at the probe in (a) the forward and (b) the backward configurations. The left column shows the sign of the (purely) thermoelectric contributions when $\mu_p=\mu$. In the right column, these contributions have modified $\mu_p$, resulting in the reversal of the flows through one of the barriers. (c) Electrochemical potential and temperature developed at the thermalization probe in the forward and backward configurations. Parameters: $\Gamma = 3\kBT$, $\mu=40\kBT$, and $\Delta T/T=1$.}
\end{figure}

The thermalization-induced bipolar thermoelectric diode occurs around the condition when the resonances of scatterers 1 and 3 are antisymmetric with respect to the base electrochemical potential, $\mu$, and have a similar (but opposite) response to electrochemical and temperature differences with the probe terminal. That is the case represented in Fig.~\ref{fig:bipolarTh}. In order to understand this effect, it is important to first notice that the temperature difference is the only thermodynamic force. Hence, the flow of heat is well defined by the second law of thermodynamics (and by everyone's intuition) and goes from the hot to the cold reservoir, i.e., it has opposite sign in the forward and the backward configurations. Differently, the sign of the thermoelectric current is not fixed by any thermodynamic law. In this particular case, it depends on the developed potential in the probe. 

Consider first the forward configuration, cf. Fig.~\ref{fig:bipolarTh}(a), in which the thermalization probe initially has an electrochemical potential $\mu$ but its temperature is already the one fixed by the probe conditions $I_p=J_p=0$. Clearly, $T_R<T_p<T_L$, see Fig.~\ref{fig:bipolarTh}(c). For concreteness, we fix $\varepsilon_1=-\varepsilon_3>\mu$. In such a situation, the thermoelectric response of the two resonant tunneling barriers results in charge flowing into the conductor: particles flow from the hotter to the colder reservoir when the resonance is over the electrochemical potential, and in the opposite direction when it is below~\cite{benenti:2017}. 
The probe terminal hence increases its electrochemical potential, as shown in Fig.~\ref{fig:bipolarTh}(c). 
The voltage established between terminals L and R and the probe introduces another contribution to the particle current (additional to the temperature-driven thermoelectric response). In this particular configuration, the distance $\varepsilon_1-\mu_p$ is reduced, while $\varepsilon_3-\mu_p$ increases, so transport through barrier 1 is most sensitive to this change. Eventually (when $\mu_p-\mu$ attains the thermovoltage of barrier 1 under a temperature difference $T_L-T_p$), the voltage-induced contribution developed by thermalization is able to reverse the flow of particles through barrier 1. The particle current through the conductor then flows from R to L and $I^F<0$.

In the backward configuration, see Fig.~\ref{fig:bipolarTh}(b), the same arguments apply, with the difference that, as the resonances are opposite, when exchanging the temperatures $T_L$ and $T_R$, the initial thermoelectric contributions have the opposite sign, i.e., the flow into the $L$ and $R$ terminals. 
Then, the electrochemical potential of the probe tends to diminish, , see Fig.~\ref{fig:bipolarTh}(c), again approaching the resonance coupled to the hot reservoir ($\varepsilon_3$), so it is the contribution through barrier 3 the one that is finally reversed. As a consequence, we again have particles flowing from $R$ to $L$, i.e., $I^B<0$. 

In the perfectly antisymmetric case, we furthermore have $I^F=I^B<0$.

\begin{figure}[t]
    \includegraphics[width=\linewidth]{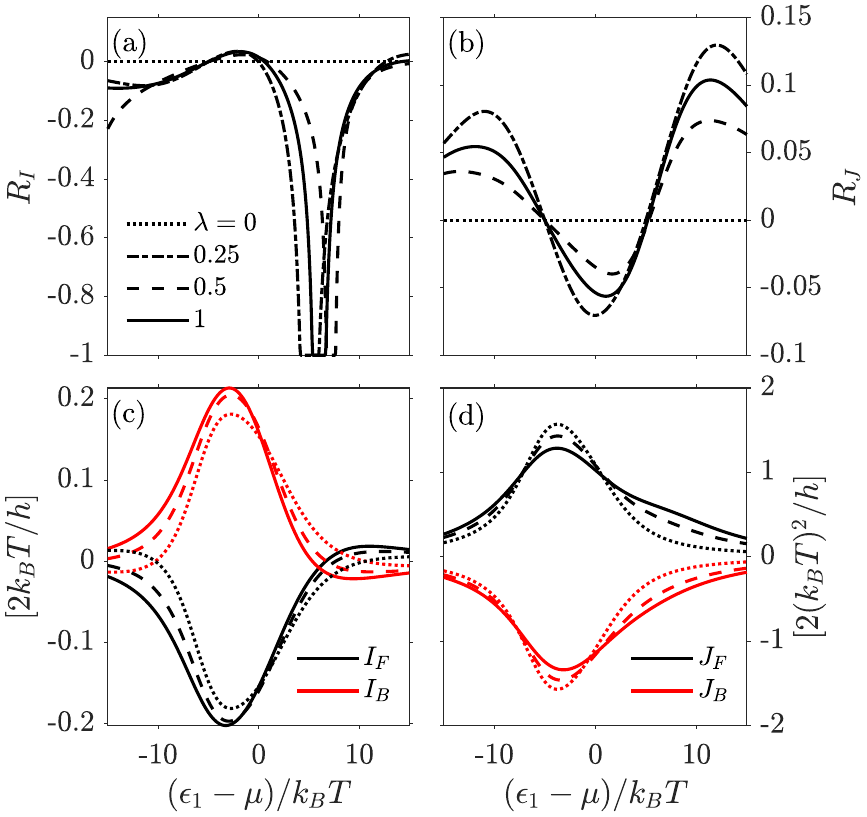}    \caption{\label{Sfig:finitelambda}Effect of the coupling to the thermalization probe on (a) the thermoelectric and (b) thermal rectification coefficients. (c) and (d) show the corresponding particle and heat currents. Parameters: $\Gamma = 3k_BT$, $\mu=40k_BT$, $\Delta T=T$, $d=2h/\sqrt{8mk_BT}$ and $(\epsilon_3-\mu)=-5k_BT$.
    }
\end{figure}
\subsection{Effect of the finite coupling to the probe}
\label{Ssec:finitelambda}

In the main text we have discussed the case where all the electrons entering the center region are thermalized by the probe, with $\lambda=1$. The effect of a finite coupling to the probe is plotted in Fig.~\ref{Sfig:finitelambda}, showing that the rectification coefficients and currents have the same qualitative behaviour. In particular, the bipolar thermoelectric effect is robust to finite $\lambda$, though the region where it appears is reduced with the opacity of the coupling. 

\subsection{Temperature dependence}
\begin{figure}[t]
    \includegraphics[width=\linewidth]{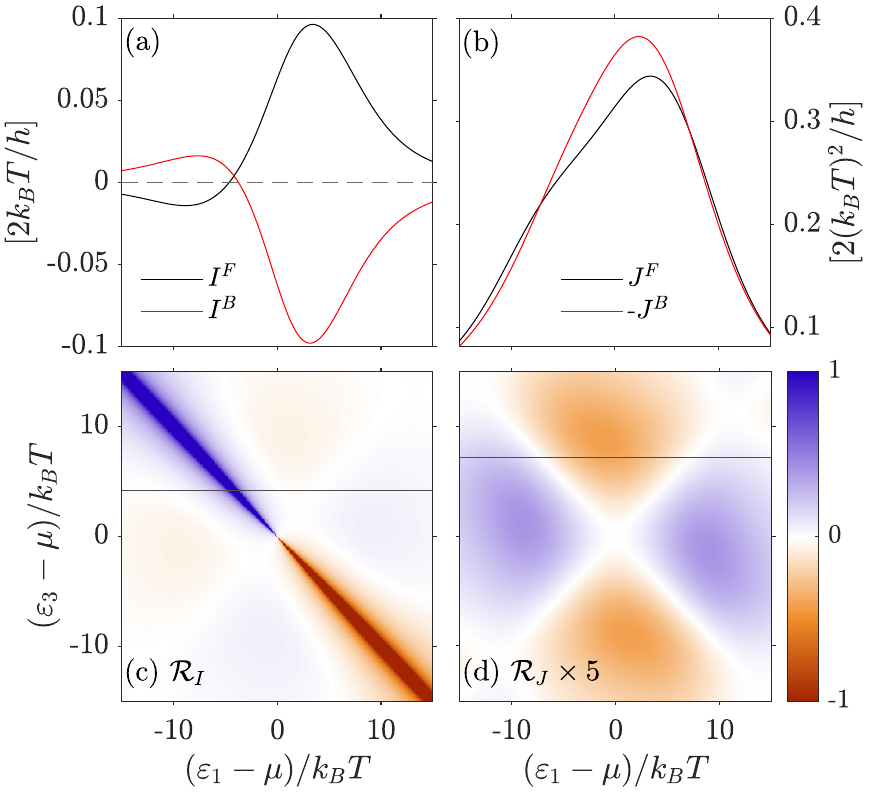}
    \caption{\label{Sfig:ThDT05}Thermalization diode. Forward and backward (a) particle and (b) heat currents as functions of $\varepsilon_1$ tuned along $(\varepsilon_3{-}\mu)/\kBT=4.8$ and $8.7$, respectively, indicated by black lines  in (c) and (d), correspondingly. The later panels show the thermoelectric and heat rectification coefficients, as functions of both resonances. The thermalization probe is fully coupled ($\lambda=1$), and $\mu\gg U_\alpha$, so the band bottom has no effect. Parameters: $\Gamma = 3\kBT$, $\mu=40\kBT$, and $\Delta T/T=1/2$. }
\end{figure}

We plot in Fig.~\ref{Sfig:ThDT05} the currents and rectification coefficient for $\Delta T=T/2$, showing that, though expectedly weaker, the same effects discussed in the main text for $\Delta T=T$ also appear in this case. Antireciprocal thermoelectric currents are restricted to a narrower region around $\varepsilon_1=-\varepsilon_3$. This is because the shift of $\mu_2$ is not large enough to invert one of the currents far from the antisymmetric condition.

\begin{figure}[t]
    \includegraphics[width=\linewidth]{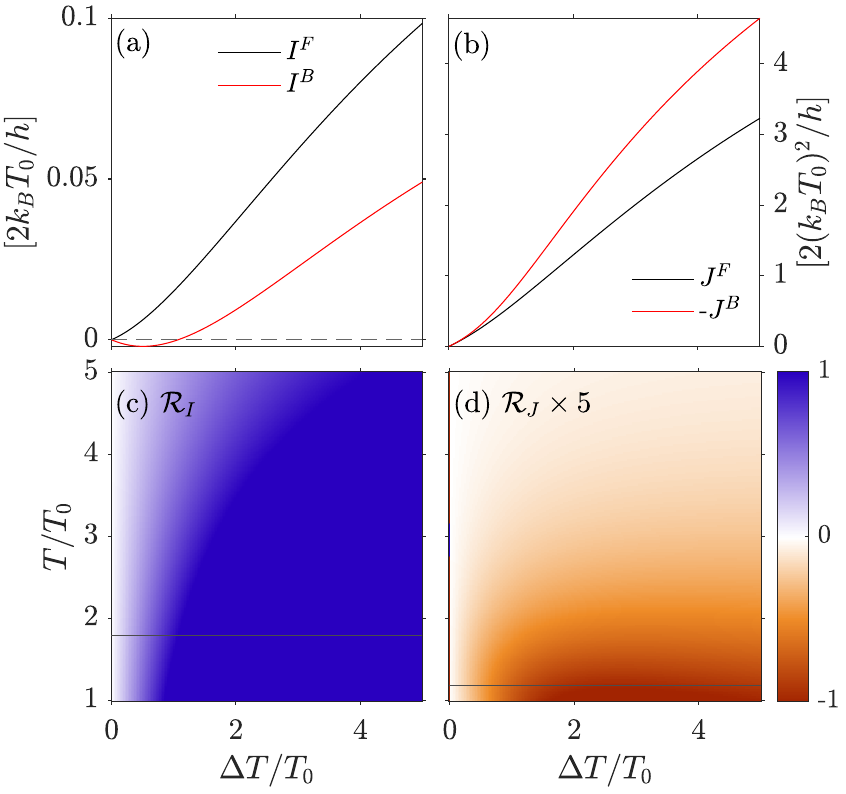}    \caption{\label{Sfig:Th_Tdep}Temperature dependence of the thermalization-induced rectification. Forward and backward (a) particle and (b) heat currents as functions of $\Delta T$ for temperatures $T=1.8T_0$ and $1.2T_0$, respectively, indicated by black lines in (c) and (d), correspondingly. The later panels show the thermoelectric and heat rectification coefficients, as functions of $\Delta T/T_0$ and $T/T_0$, for configurations of Fig. 3 of the main text with $(\epsilon_1-\mu)/k_BT_0= -6$ and $0$, and $(\epsilon_1-\mu)/k_BT_0= 7$ and $10$ respectively.
    The remaining parameters are the same, with $\lambda=1$. }
\end{figure}

Figure~\ref{Sfig:Th_Tdep} shows the temperature dependence of the thermalization-induced bipolar thermoelectric and thermal diode effects for two configurations of Fig. 3 of the main text. The one for the thermoelectric case is chosen slightly de-tuned from the antisymmetric configuration with $\varepsilon_1+\varepsilon_3=0$, for which we find $\mathcal{R}_I=1$ for every $\Delta T$, $T$. In this particular configuration, the bipolar thermoelectric diode still occurs for almost all temperature configurations, see Figs.~\ref{Sfig:Th_Tdep}(a) and \ref{Sfig:Th_Tdep}(c). Only close to the linear response regime, with small temperature differences, we have ${\cal R}_I<1$. As shown in Figs.~\ref{Sfig:Th_Tdep}(b) and \ref{Sfig:Th_Tdep}(d), the difference of the forward and backward currents increases with $\Delta T$, as expected. The thermal rectification coefficient however saturates. We have introduced a new reference temperature $T_0$ specifically for this plot. 

\section{Bipolar diode by screening effects}
\label{Ssec:bipolarRect_screening}

\begin{figure}[t]
    \includegraphics[width=\linewidth]{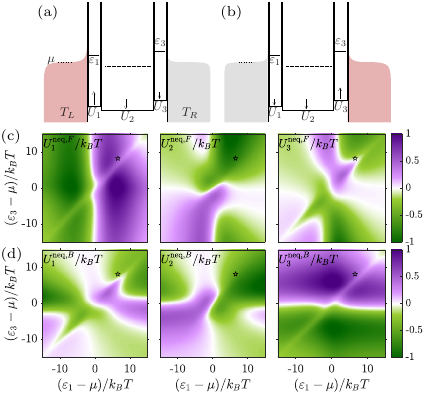}    \caption{\label{Sfig:screensqueme}Scheme of the developed internal potentials, $U_\alpha$,  induced by charge accumulation in (a) the forward and (b) the backward configurations. Solid lines mark the energy of the resonances, $\varepsilon_1<\varepsilon_3$, with the dotted line indicating the position of a Fabry-Perot interference peak. Under the appropriate conditions, this effect results in a bipolar thermoelectric diode with particles flowing from $L$ to $R$ in both configurations.
    Developed nonequilibrium electrostatic energies in the three regions for the (c) forward and (d) backward configurations. Same configuration as in Fig. 3 in the main text, with $\lambda=0$, $d=2 h/\sqrt{8mk_BT}$, $\Gamma = 1\kBT$, $\mu=40\kBT$ and $\Delta T=T$.
    }
\end{figure}
\begin{figure}[t]
    \includegraphics[width=\linewidth]{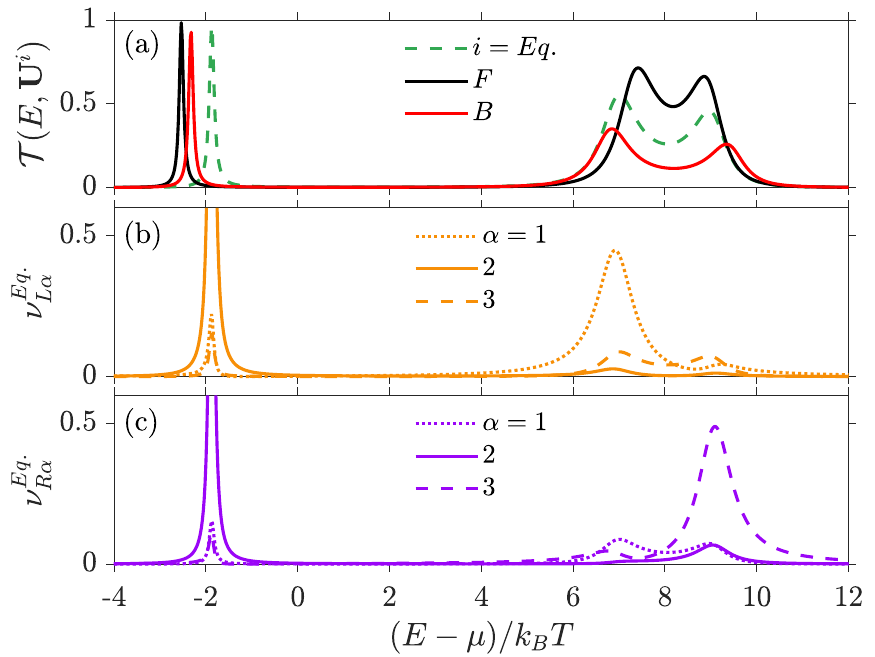}    \caption{\label{Sfig:NL_transmissions}(a) Transmission probability when the system is in equilibrium (solid green), and in the forward (dashed black) and backward (dotted red) configurations, once the electrostatic potentials have build-up in the three areas as a result of the corresponding non-equilibrium situation. (b)-(c) Injectivities in regions $\alpha=1,2,3$ from reservoirs $L$ and $R$, respectively. The parameters correspond to the ones where $I^B=I^F\neq0$ in Fig. 3(e) of the main text, marked by $\star$ in Fig.~3(a) [as well in Fig.~\ref{Sfig:screensqueme}(c) and \ref{Sfig:screensqueme}(d)].
    }
\end{figure}

The screening-induced bipolar thermoelectric diode occurs when the transmission energy dependence presents two nonzero contributions, one above and one below the reference electrochemical potential, as sketched in Fig.~\ref{Sfig:screensqueme} and plotted in Fig.~\ref{Sfig:NL_transmissions}(a): a wide double-peak at positive energies, due to the two Breit-Wigner resonances being close in energy, and a sharp peak at negative energies, due to the Fabry-Perot interference in region 2. The nonlinear response will depend on how the internal energies $U_\alpha$ react to the temperature increase in one of the reservoirs, therefore modifying the overall transmission probability, ${\cal T}(\{\mathbf{U}\})$. 
The internal energies are obtained self-consistently by solving the change in the induced charge, $\delta q_\alpha$, in each region. This is given by the injectivities $\nu_{l\alpha}$ from terminal $j$ into region $\alpha$ via
\begin{equation}
\label{Seq:dq_q}
   \frac{\delta q_\alpha}{-e} = {\int}dE \sum_j  [\nu_{j\alpha}(E,\mathbf{U})f_j(E){-}\nu_{j\alpha}^{\rm eq}f_{\rm eq}(E)],
\end{equation}
see main text.
Clearly for a given region the injectivity is larger for the terminal with which it is directly coupled i.e., $\nu_{L1}\gg\nu_{R1}$ and $\nu_{R3}\gg\nu_{L3}$, see Figs.~\ref{Sfig:NL_transmissions}(b) and \ref{Sfig:NL_transmissions}(c). Hence, one expects the potential of region 1(3) to change more in the forward (backward) configuration. The opposite region in each configuration is expected to be barely affected by the temperature increase, therefore this small modification can then also be influenced by the charging of the other regions via the self-consistency.  

To get a better intuition, let us focus on a particular configuration, as the one sketched in Fig.~\ref{Sfig:screensqueme}, with both Breit-Wigner resonances having energies over the chemical potential, $\mu<\varepsilon_1<\varepsilon_3$. Close to equilibrium (for small temperature differences), the current is dominated by the Breit-Wigner resonances, so particles flow from L to R in the forward, and opposite in the backward configuration. Consider first the forward configuration, depicted in Fig.~\ref{Sfig:screensqueme}(a). 
The change in the nonequilibrium electrostatic energies is shown in Fig.~\ref{Sfig:screensqueme}(c) as the resonances are tuned.
The temperature increase in terminal L contributes to charge region 1 by increasing the occupation of the lead at the energies around $\varepsilon_1$ (for enhancing the tail of the Fermi distribution). Hence, $U_1$ is expected to increase. The next region to consider is region 2, whose coupling to terminal L is weaker. 
This region features a sharp interference peak below the chemical potential, thus resulting on a decrease of the charge with respect to the equilibrium state (states of the lead below the chemical potential are less occupied when increasing the terminal temperature), so $U_2$ decreases.  
Finally, region 3 is in this case little affected by the temperature variation in terminal L, but it can be indirectly affected by the potential variation in the regions in between: the change in $U_1$ and $U_2$ also affects region 3 and compete in the resulting $U_3$ is a complicated manner. In this particular case, it also decreases. 
%further reduces the overlap of states in region 3 with the {\it distant} terminal L, resulting in a reduction of the induced charge, so $U_3$ tends to decrease. 
With this, the two Breit-Wigner resonances in regions 1 and 3 are then brought closer in energy, resulting in a narrower and sharper feature at positive energies in the transmission probability, see Fig.~\ref{Sfig:NL_transmissions}(a),  which is favorable for the thermoelectric current from L to R. 

The same arguments apply to the backward configuration, depicted in Fig.~\ref{Sfig:screensqueme}(b). The developed $U_\alpha^{\rm neq}$ are shown in Fig.~\ref{Sfig:screensqueme}(d). In this case, it results in $U_1$ decreasing and $U_3$ increasing, which separates the two Breit-Wigner resonances and makes the double peak at positive energies wider and lower, see Fig.~\ref{Sfig:NL_transmissions}(a). Thus the thermoelectric contribution at positive energies (from the hot R to the cold L) is reduced. The Fabry-Perot interference peak, being at negative energies contributes to transport in the opposite direction (from L to R). Eventually, the Breit-Wigner contribution gets so weak that the Fabry-Perot contribution, which is robust to the change in temperature, starts to dominate. At that point, the current through the system changes sign, so particles flow from L to R, as in the forward configuration.  
%\sout{In the absence of screening, features of energies above $\mu$ contribute with electrons flowing from the hot reservoir to the cold one. Conversely, features of energies below $\mu$ contribute to transport from cold to hot. Screening effects are able to modify the direction of transport with respect to this scenario by changing the relative dominance between these. }

In other words, the antireciprocal effect is due to a change of the contribution of positive- and negative-energy spectral features, which contribute oppositely to the thermoelectric response.

\subsection{Temperature dependence}
\label{Ssec:NL_Tdep}
\begin{figure}[t]
    \includegraphics[width=\linewidth]{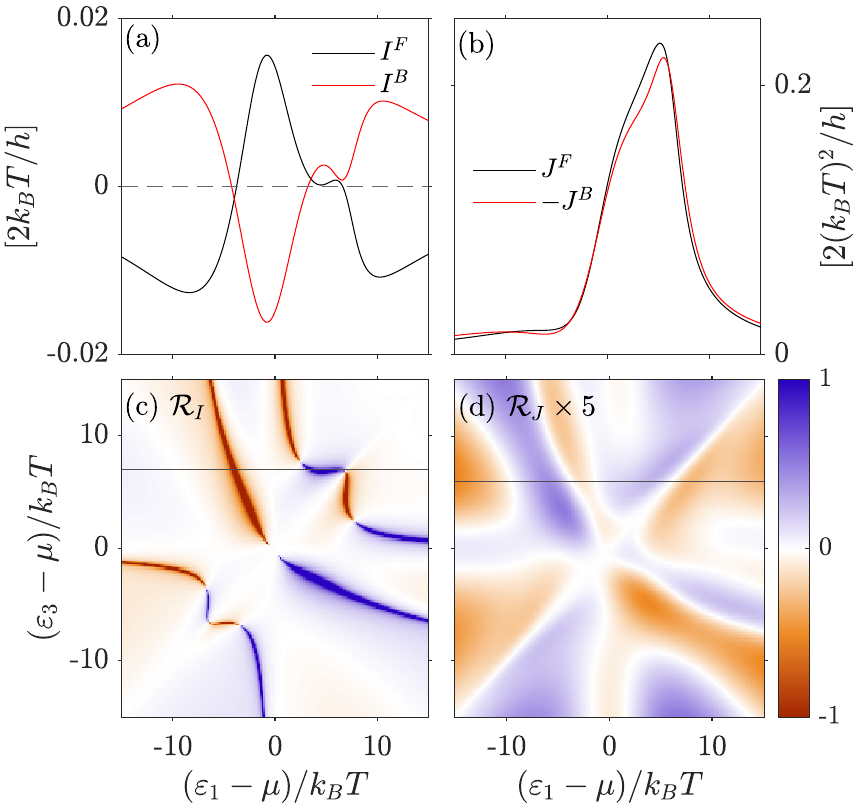}
    \caption{\label{Sfig:NL_DT05}Rectification by screening. Forward and backward (a) particle and (b) heat currents as functions of $\varepsilon_1$, for fixed $(\varepsilon_3-\mu)/k_bT=8.2$ and $7.5$ as marked by black lines in (c) and (d), respectively. The later show the dependence of ${\cal R}_I$ and ${\cal R}_J$ with gating the RTBs. Parameters: $\lambda{=}0$, $d=2 h/\sqrt{8mk_BT}$, $\Gamma = \kBT$, $\mu=40\kBT$, $\Delta T/T=1/2$. }
\end{figure}

Figure~\ref{Sfig:NL_DT05} shows the thermoelectric and thermal rectification properties at a lower temperature difference than the one considered in the main text, $\Delta T=T/2$. The same features observed in Fig. 3 of the main text appear, with smaller rectification coefficients, as expected. The bipolar thermoelectric diode occurs in narrower regions of the gate-voltage map.

\begin{figure}[t]
    \includegraphics[width=\linewidth]{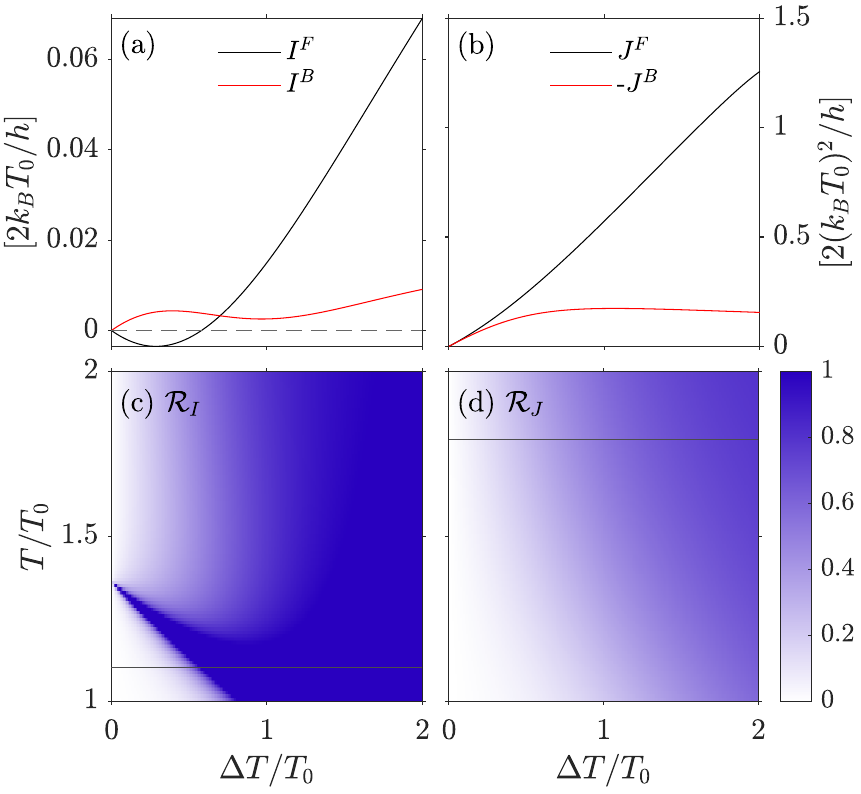}    \caption{\label{Sfig:NL_Tdep}Temperature dependence of the screening induced rectification. Forward and backward (a) particle and (b) heat currents as functions of $\Delta T$ for temperatures $T=1.1T_0$ and $1.8T_0$, respectively, indicated by black lines in (c) and (d), correspondingly. The later panels show the
    thermoelectric and heat rectification coefficients, as functions
    of $\Delta T/T_0$ and $T/T_0$. The remaining parameters correspond to the ones where $I_B=I_F\neq0$ in Fig. 3(e) of the main text, marked by $\star$ in Fig.~3(a) [as well in Fig.~\ref{Sfig:screensqueme}(c) and \ref{Sfig:screensqueme}(d)].
    }
\end{figure}

The temperature dependence is plotted in Fig.~\ref{Sfig:NL_Tdep} for the particular gating corresponding to Fig.~\ref{Sfig:NL_transmissions} (marked with a $\star$ in Fig.~3(a) of the main text). It shows that the bipolar thermoelectric diode occurs for wide temperature ranges, cf. Figs.~\ref{Sfig:NL_Tdep}(a) and \ref{Sfig:NL_Tdep}(c), showing a threshold at low temperatures and temperature gradients which is tunable with gating. The thermal currents show a robust diode effect, with the $I^F$ increasing with $\Delta T$, while $I^B$ saturates and even shows negative thermal differential conductance for large enough $\Delta T$, cf. Fig.~\ref{Sfig:NL_Tdep}(b). The thermal rectification coefficient increases both with $T$ and $\Delta T$, see Fig.~\ref{Sfig:NL_Tdep}(d).

%%%%%%%%%%%%%%%%%%%%%%%%%%%%%%%%%%%%%%%%%%%%%%%%%%%%%%%%%%%%%%%%%%%%%%%%%%%%%%

%%%%%%%%%%%%%%%%%%%%%%%%%%%%%%%%%%%%%%%%%%%%%%%%%%%%%%%%%%%%%%%%%%%%%%%%%%%%%%
%%%%%%%%%%%%%%%%%%%%%%%%%%%%%%%%%%%%%%%

\section{Charge and heat currents}
\label{Ssec:currents}

\begin{figure*}[t]
    \includegraphics[width=\linewidth]{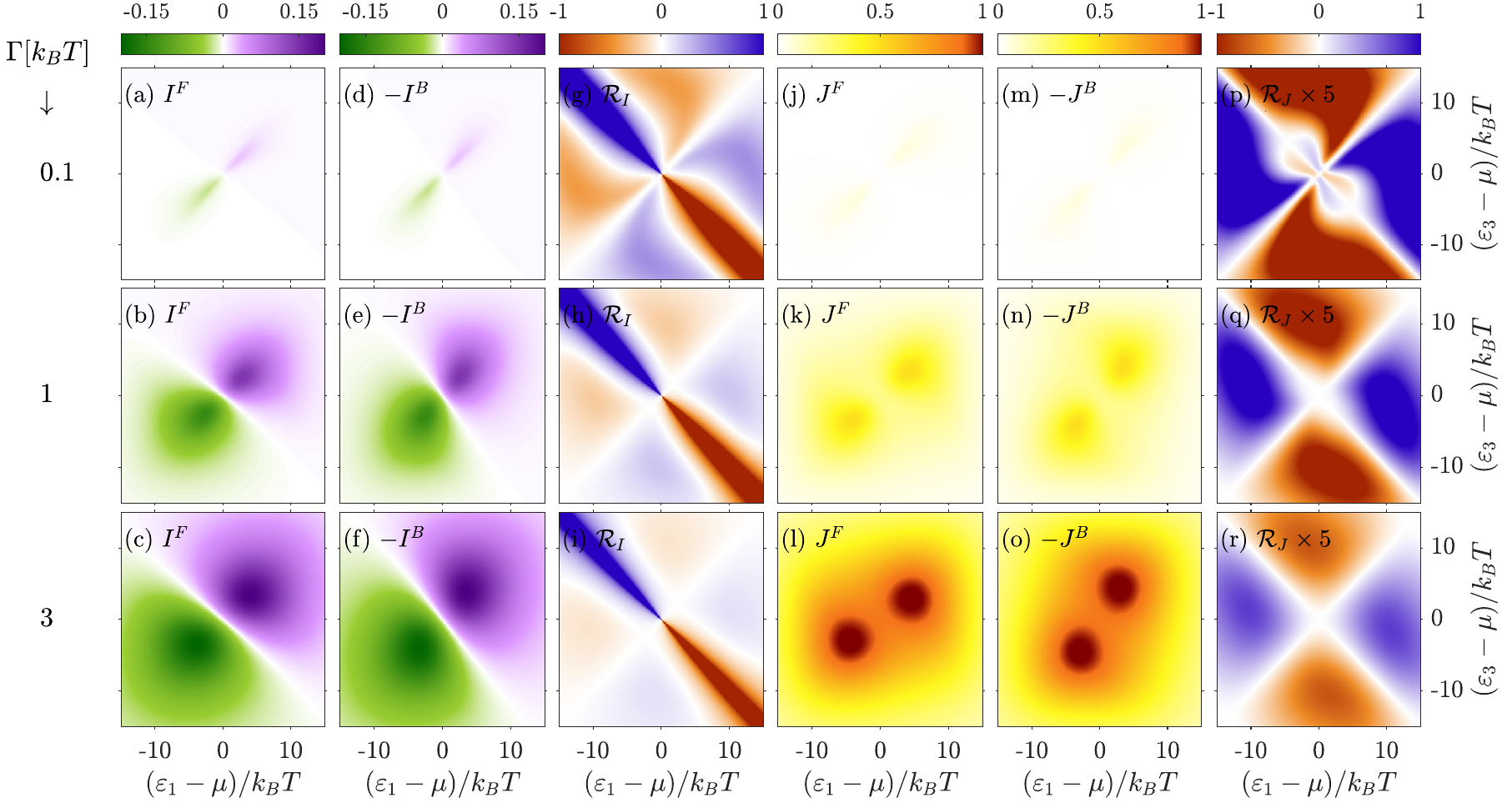}
    \caption{(a)-(c) Forward and (d)-(f) backward particle currents, and (g)-(i) thermoelectric rectification for the double resonant tunneling barrier conductor sandwiching a strong thermalization region ($\lambda=1$) and neglecting screening, as functions of the resonance energies, $\varepsilon_1$ and $\varepsilon_3$ and for increasing quantum dot linewidth, $\Gamma$. (j)-(l), (m)-(o) and (p)-(r) show the corresponding the heat quantities. In all cases, $\Delta T=T$.}
    \label{Sfig:therm}
\end{figure*}

\begin{figure*}[t]
    \includegraphics[width=\linewidth]{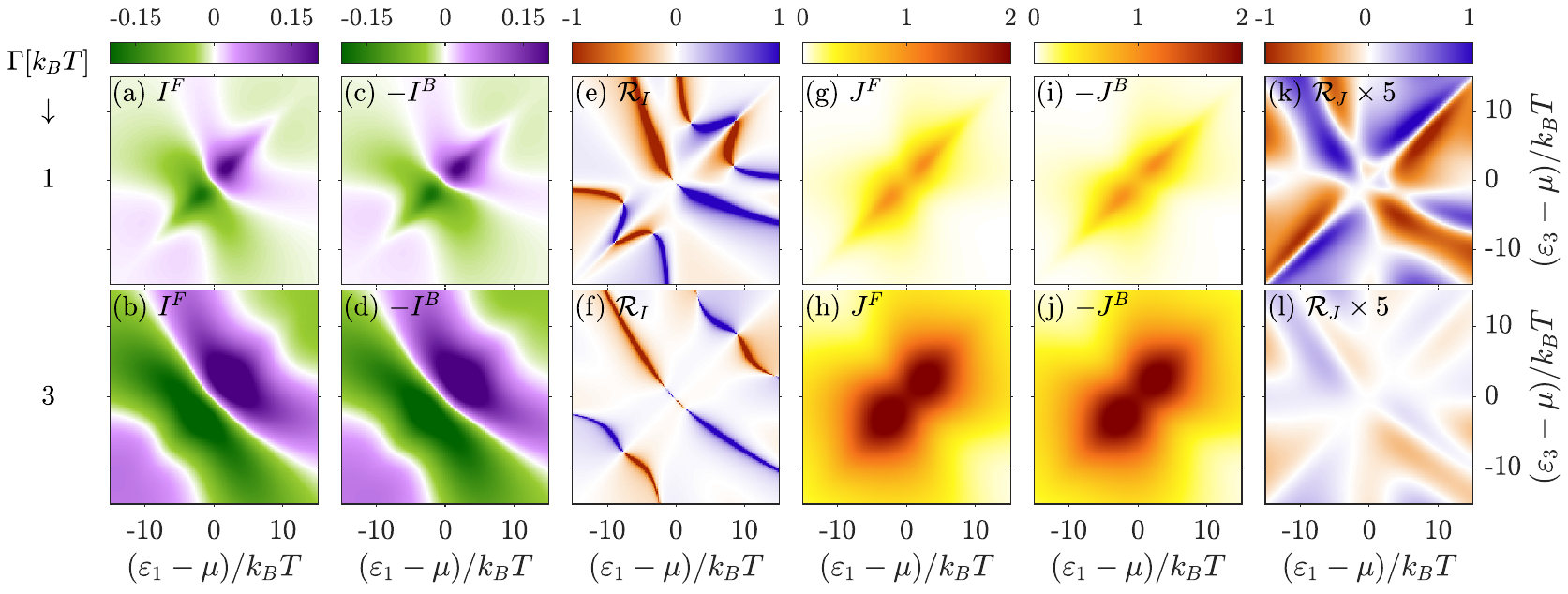}
    \caption{Same as Fig.~\ref{Sfig:therm} in the elastic case with $\lambda=0$ including screening effects, for $d=2h/\sqrt{8mk_BT}$ and $\mu=40k_BT$.}
    \label{Sfig:scr}
\end{figure*}

\begin{figure*}[t]
    \includegraphics[width=\linewidth]{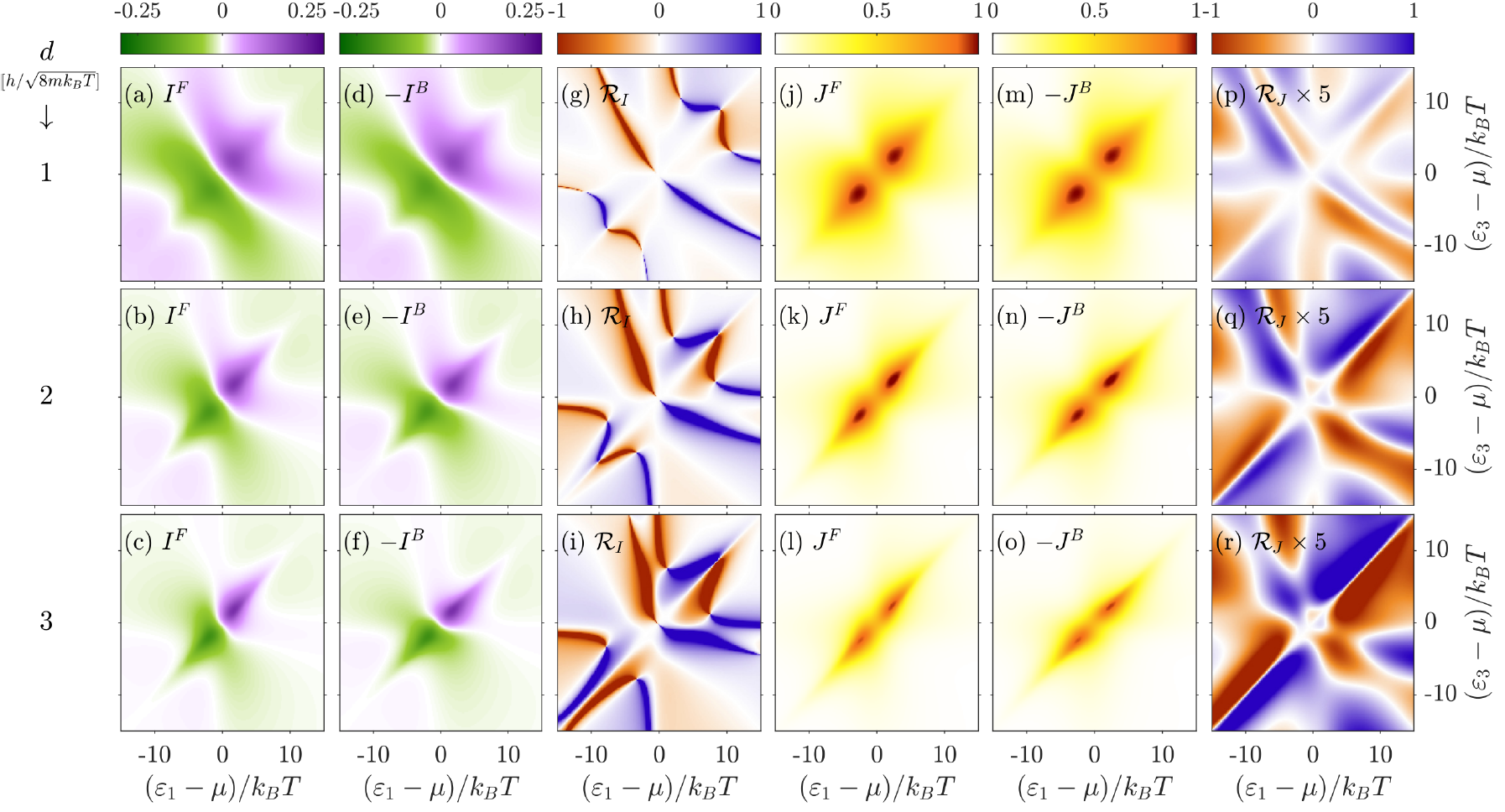}
    \caption{Same as Fig.~\ref{Sfig:scr} for $\Gamma=\kBT$ and for different lengths of region 2, $d$, which controls the position and width of the Fabry-Perot resonance at energies below the chemical potential.}
    \label{Sfig:scr_d}
\end{figure*}
\begin{figure*}[t]
    \includegraphics[width=\linewidth]{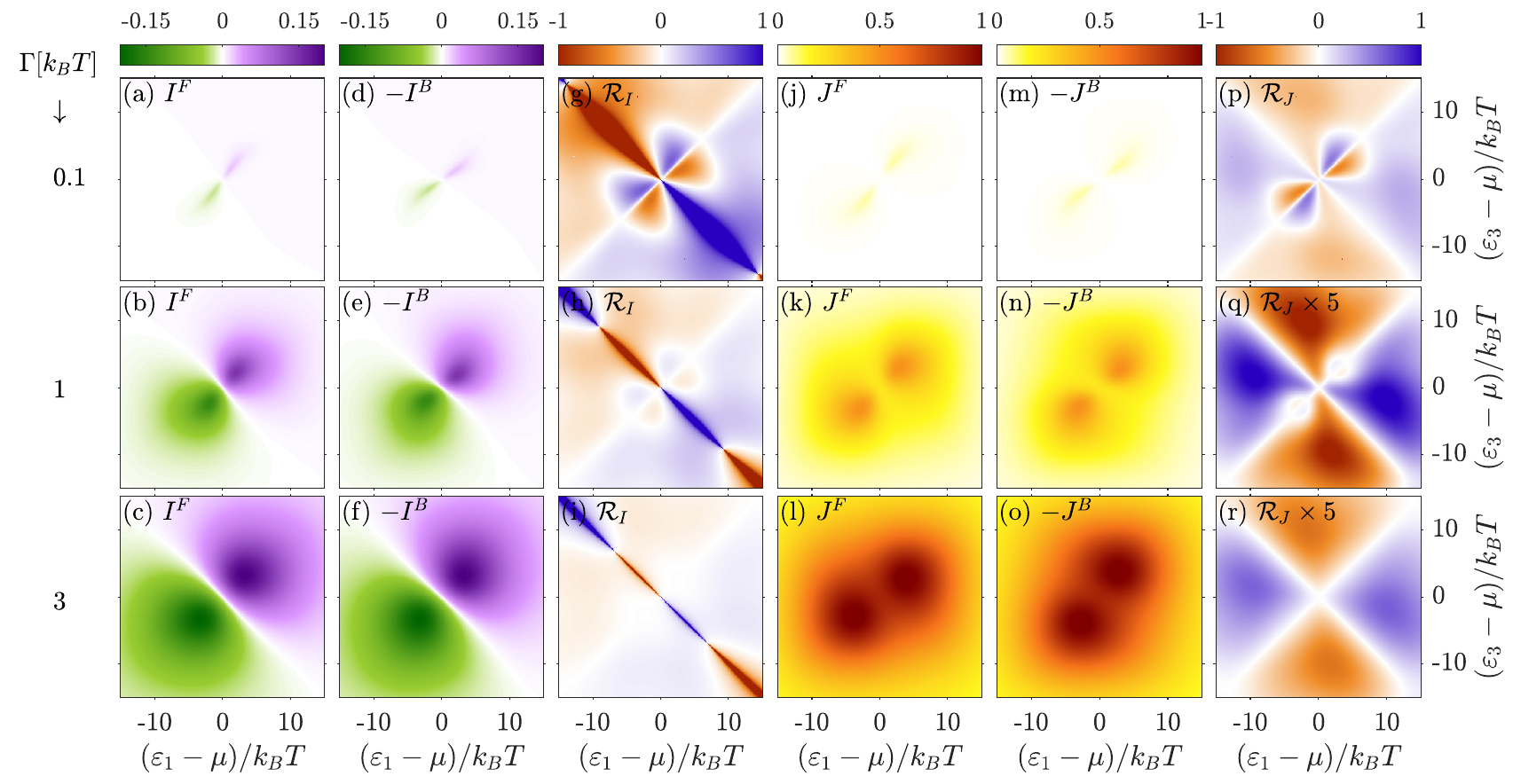}
    \caption{Same as Fig.~\ref{Sfig:scr} in the strong thermalization regime with $\lambda=1$.}
    \label{Sfig:mix}
\end{figure*}

Figures~\ref{Sfig:therm}, \ref{Sfig:scr}, \ref{Sfig:scr_d} and \ref{Sfig:mix} show the forward and backward particle and heat currents as well as the thermoelectric and thermal rectification coefficients for the different configurations as functions of the quantum dot linewidth, $\Gamma$, and of the distance between barriers 1 and 3, $d$ (for the thermalization free case with $\lambda=0$, Fig.~\ref{Sfig:scr_d}). 
The thermoelectric currents reach values of the order of \unit{nA} at $T=\unit[1]{K}$ and $\Delta T/T=1$, where a bidimensional GaAs device would have a carrier concentration of $\sim\unit[10^{12}]{cm}^{-2}$ and sizes of tens of \unit{nm} for the explored regimes.

\end{document}